\documentclass[11pt,twoside,A4]{article}
\usepackage{listings}
\usepackage{times,fancyhdr}
\usepackage[dvips]{graphicx}
\usepackage{latexsym}
\usepackage[affil-it]{authblk}
\usepackage{mathptm}
\usepackage{amsmath}
\usepackage{amssymb}
\usepackage{color}
\usepackage{setspace}
\usepackage{hyperref}
\usepackage{todonotes}
\usepackage[top=4cm, bottom=4cm, left=4cm, right=4cm]{geometry}
\begin{document}

\title{Predicting Citation Counts with a Neural Network}
\author{Tobias Mistele${}^*$, Tom Price, Sabine Hossenfelder${}^*$}
\affil{\small ${}^*$ Frankfurt Institute for Advanced Studies\\
Ruth-Moufang-Str. 1,
D-60438 Frankfurt am Main, Germany
}
\date{}
\maketitle
\begin{abstract}
We here describe and present results of a simple neural network that predicts
individual researchers' future citation counts based on a variety of data from
the researchers' past. For publications available
on the open access-server \href{http://arxiv.org}{arXiv.org} we find a higher predictability than previous studies.
\end{abstract}

\section{Introduction}
Measuring and predicting scientific excellence is as daunting as controversial. But regardless of how
one feels about quantifying quality, measures for
scientific success are being used, and they will continue to be used. The best that we, as scientists,
can do is to at least come up with good measures.

The attempt to quantify scientific success by the number of publications dates back to the
early 19th century \cite{Nature2017}, but the idea really took off with the advent
of the internet when data
for publications and citations became easily accessible. Since then,
a large variety of different measures has been suggested which rely on citation counts, publication
statistics, impact factors, media coverage, and more.  For a recent review, the reader is
referred to \cite{Review2015}.

We maintain that the best way to assess a researcher's promise is to study their work
in depth. But bowing to the need to have a tool for administrative and
organizational purposes that is fast and easy to use and allows at least superficial evaluation, we here report on
the training of a neural network to
improve on presently existing measures.

Due to the large variety of existing measures, we will not compare the method presented here
to all of them. We will focus in particular on three previous studies that
put our work into perspective. This is (1) the original paper about the predictivity
of the Hirsch-index \cite{Hirsch2007} (hereafter $h$-index), (2) a 2012 study \cite{Acuna2012}
which used a linear regression model to predict the $h$-index for research in the life sciences,
and (3) a 2017 report \cite{7991559} on various machine learning models used to predict the $h$-index
for a large cohort of authors in the computer sciences.

Our paper is organized as follows. In the next section we clarify exactly
what we aim to achieve. In section \ref{data} we document which input
data we have used and how we have encoded it. In section \ref{nn} we
explain how we set up the neural network, and in \ref{results} we will
present our results. We finish with a discussion and conclusion in section
\ref{conc}.

\section{Aim}
\label{aim}

Before we build a neural network, we first need to make precise what we mean by ``predictive'' and how we will measure this ``predictiveness.''

Our neural network will be fed publication data (section \ref{data}) for a training group of researchers during a first phase of publishing activity,
hereafter referred to as the `initial period'. The
aim is then to use the neural network (section \ref{nn}) to predict individual authors' performance in a second phase of publishing activity, hereafter referred to as the `forecasting period'. This input data for the initial period does not include citations from papers that were only published during the forecasting period. The prediction period starts at 1/1/2008 which is chosen such that we can evaluate the neural network's performance for predictions up to 10 years into the future of the initial period.

Once the network has learned forecasting from the training group, we make a forecast for the remaining researchers -- the ``validation group'' -- and evaluate
how good our forecast was. This is to say, in this present study we do not make actual future forecasts because we want to assess how well out network performs, but our method is designed so that it could be used to make real forecasts.

We will use two different approaches to evaluate the forecasting performance in the validation group.

The first approach (see section \ref{first}) follows the procedure used in Ref.~\cite{Hirsch2007}, which evaluated the predictiveness of the $h$-index for various quantities in terms of the correlation coefficient, $ r $. In this approach, we do not include citations from papers in the initial period in the number of citations to be forecasted in the forecasting period.
In making this clean separation,
we get a better grasp on predicting \textit{future} scientific achievement as opposed to \textit{cumulative} achievement.

For this first approach, we follow the notation of \cite{Hirsch2007} and denote the number of citations received between $t_1$ and $t_2$, i.e. during the
forecasting period, as $ N_\textrm{c}(t_1, t_2)$. Since it is known that the $h$-index roughly
scales with the square-root of citations, we will more precisely use $ N_\textrm{c}(t_1, t_2)^{1/2}$ to
make it easier to compare our results with those of Ref.~\cite{Hirsch2007}.

The second approach (see section \ref{second}) follows a different procedure which is better suited for comparison with
the results of Refs.~\cite{Acuna2012,7991559}.
For this, we feed the neural network the same input data as in the first approach but then predict the cumulative $ h $-index after $ n $ years until the end of the forecasting period.
Furthermore, in this second approach we use the coefficient of determination, $ R^2 $, instead of the correlation coefficient, $ r $, to quantify the goodness of our prediction because the same procedure was followed in Refs.~\cite{Acuna2012,7991559}.

Unless otherwise stated, our general procedure is to employ 20 rounds of Monte Carlo cross-validation -- i.e. redo the random split into training and validation data 20 times and retrain the neural network -- and report the mean as well as the standard deviation.
One reason for employing cross-validation is that different splits of our dataset into training and validation data leads to slightly different results as will be further discussed below.
Another reason is that our dataset is not particularly large. Cross-validation then allows to avoid splitting our dataset into a training, a validation, and test group while still avoiding overfitting to a particular split into training and validation data.

\section{Data}
\label{data}

We have obtained the publications for each author from the arXiv through the publicly available Open Archives Initiative API \cite{arXivOAI}
and corresponding citation data from Paperscape \cite{paperscape}.
For the purposes of this present work, we consider only the publications in the  `physics' set, which gives us a total of
934,650 papers.
Journal Impact Factors ({\sc JIF}s) are taken from Ref.~\cite{JCR2017}.
We group together similar author names and treat them as a single author by the same procedure as laid out in \cite{Tom}.

From the complete dataset, we select a sample of authors and trim it in various ways. First, we require that they published their first arXiv paper between 1/1/1996 and 1/1/2003.
We have chosen that period to span the time between 5 and 12 years prior to the cutoff which matches the procedure of Ref.~\cite{7991559}.

We then remove authors who have published fewer than 5 or more than 500 papers to avoid statistical outliers which would unduly
decrease the predictivity of our method.

Finally, we exclude collaborations, since their publication activity differs greatly from that of individuals. For
this, we remove all author names that contain the word  `collaboration' and all papers with
more than 30 authors.

After this, we are left with a sample of 39,412 author IDs. From these, we randomly chose a subset of 28,000
as the `training group'. The rest is our validation data by help of which we evaluate how
well the neural network performs after training is completed.
Note that this random split into training and validation data is done independently for each round of cross-validation.

\section{The Neural Network}
\label{nn}

The neural network itself is built using Keras \cite{keras} with the TensorFlow backend \cite{tensorflow}.

We used a feedforward neural network, which means that the neural network consists of layers of neurons where the input of the neurons in one layer is the output of the neurons in the previous layer and the layers are not arranged as a cycle.
The output of the first layer are the input data described in Appendix A, and the output of the last layer is taken as the output of the whole neural network.
In our case, the last layer consists of ten neurons such that the neural network's output is list of ten real numbers.

For this network, the output of the neurons in one layer follows from the output of the neurons in the previous layer by
\begin{align}
 x' = \sigma\left( W \cdot x + b \right) \,.
\end{align}
Here, $ x' $ is a vector which contains the outputs of the $ N' $ neurons in one layer, $ x $ is a vector which contains the outputs of the $ N $ neurons in the previous layer, $ W $ is a real $ N' \times N $ matrix whose elements are called the weights and $ b $ is a real vector with $ N' $ elements which are called the biases.
Further, $ \sigma $ is a function which is applied to each element of the vector $ W \cdot x + b $ and is called the activation function.
The weights and biases are different for each layer, so that for each added layer one gets another weight matrix and another bias vector.
These are the free parameters of the neural network which are determined by the training procedure.

\begin{figure}[ht]
 \centering
  \includegraphics[width=.95\textwidth]{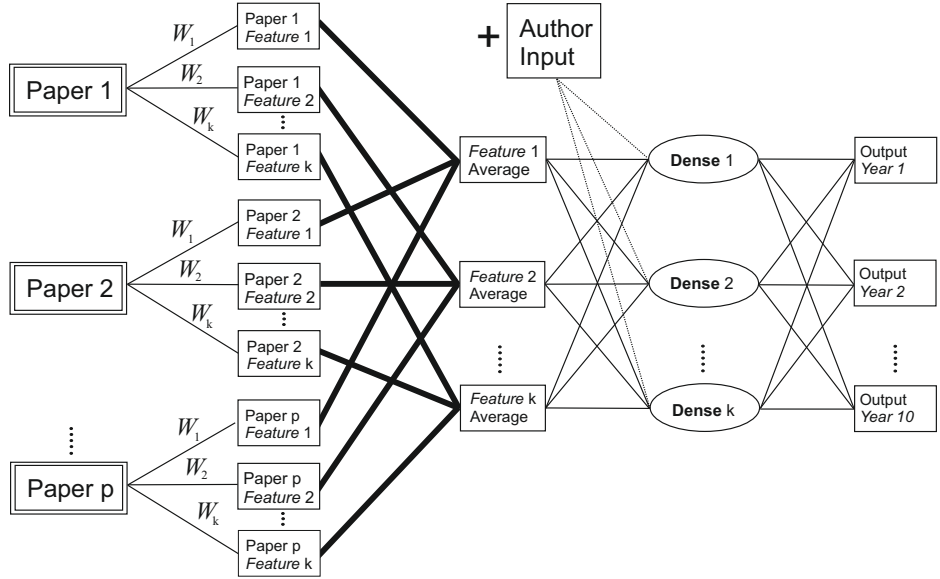}
 \vspace*{-2cm}
 \caption{Flow diagram of the neural network.  Network elements surrounded by single thin lines are single neurons; those whith double thin lines are
collections of neurons. Connections shown with bold lines have a fixed weight of $1$.}
 \label{fig:flow}
\end{figure}

During the training of the neural network, the output of the neural network is calculated with the training data described above as input and the weights and biases are adjusted in order to get the output of the neural network for each author as close as possible to the actual $ N_{\textrm{c}}(t_1, t_2)^{1/2} $.
More precisely, the weights and biases are adjusted in order to minimize the so-called loss function which we take to be the mean squared error across all authors in the training set.
Note that for a so-called fully-connected layer all elements of the weights matrix and the bias vector are independently adjusted, while for other types of layers, e.g. so-called convolutional layers, some structure is imposed. As will be explained below, we use both types of layers.

Since our dataset is not particularly large, we tried to avoid overfitting by reducing the number of parameters
and structured the network as follows (see Figure \ref{fig:flow}):

\begin{itemize}
\item The first layer is a convolutional layer and contains k = 70 neurons for each paper. Although convolutional layers are normally used to exploit translational symmetry in the input data, we use a convolutional layer with a convolution window of a single paper in order to ensure invariance under permutations of the papers. The effect of this is that each paper has a corresponding set of 70 neurons which see only the data from that paper, and each paper's neuron-set has matching weights and biases. The input to this layer contains every input except the broadness value, i.e., it contains all per paper input but not the per author input.

\item In a next step, these 70 neurons per paper are reduced to 70 neurons in total by averaging each of the 70 neurons over the different papers.
After this, no information about individual papers is left, and only average values are retained in the remaining 70 neurons.
This layer does not add free parameters to the neural network.

\item After that, a fully-connected layer with 70 neurons is added.
In addition to the output of the previous layer, this layer obtains input which is specific to the author not to
the individual papers.

\item The final layer is a fully-connected layer with ten neurons with a ReLu activation function.
The first neuron represents the prediction 1 year after the cutoff and the other neurons represent the differences between the prediction after $ n $ and $ n -1 $ years.
For instance, if the neural network's output is $ [5, 0, 1, ...] $, the corresponding prediction is $ 5 $ for $ 1/1/2009$, $ 5 $ for $ 1/1/2010$, $ 6 $ for $ 1/1/2011$, etc.
This ensures that the neural network's prediction is a monotonically increasing time series.

\end{itemize}

A detailed list of input data for each level can be found in Appendix A.
The neural network architecture described above can be implemented in Keras with just a few lines of code which
are reproduced in Appendix B.

The neurons, except for the neurons in the output layer, are taken to have $ \tanh $ activation functions.
Training is done using an Adam optimizer and a mean squared error loss function.
No regularization is employed and the final result is obtained after 150 epochs of training with a batch size of 50.

We would like to end this section with a comment on the way the neural network described above makes predictions for different years $ n $ after the cutoff.
As described above, each year $ n $ corresponds to one of the 10 neurons in the very last layer of the neural network.
An alternative would be to have 10 neural networks with only a single neuron in the output layer.
Each of the networks would then be trained to make a prediction for one particular $ n $.
One might argue that this alternative leaves more freedom for the neural networks to learn the specific requirements in making a prediction for one specific $ n $ instead of for all $ n $ at the same time.
However, it seems this is not the case in practice, since we have tried both approaches with the only difference in the neural network architecture being the number of neurons in the output layer and the resulting performance was very similar.

\section{Results}
\label{results}

For both approaches, we will compare the neural network's performance to that of a naive $ h $-index predictor which is given only the $ h $-index of an author for which a prediction is to be made.
By this naive $ h $-index predictor we mean the following:
For a given quantity to predict, i.e. the future cumulative $ h $-index or $ N_\textrm{c}(t_1, t_2)^{1/2} $, take all authors in the training set with a given $ h $-index $ h_0 $ at the time of the cutoff.
Then, calculate the arithmetic mean of the quantity to be predicted and take this mean value as a prediction for authors in the validation set given their $ h_0 $.

Note that there may be authors in the validation set with, typically high, values of $ h_0 $ that are not present in the training set.
For those authors, the prediction is determined as follows.
First, a linear polynomial is fitted to the naive $ h $-index predictor for the values of $ h_0 $ that can be calculated in the way described in the previous paragraph.
Then, the value of the fitted polynomial is taken as the prediction for the other values of $ h_0 $.

\subsection{Comparison with the $h$-index}
\label{first}

For the neural network trained to predict $ N_\textrm{c}(t_1, t_2)^{1/2} $, we are only interested in the prediction for $ n = 10 $ years after the cutoff.
Therefore, we ignore the prediction of both the neural network and the naive $ h $-index predictor for the first 9 years after the cutoff.

\begin{figure}[ht]
 \centering
 \includegraphics[width=.9\textwidth]{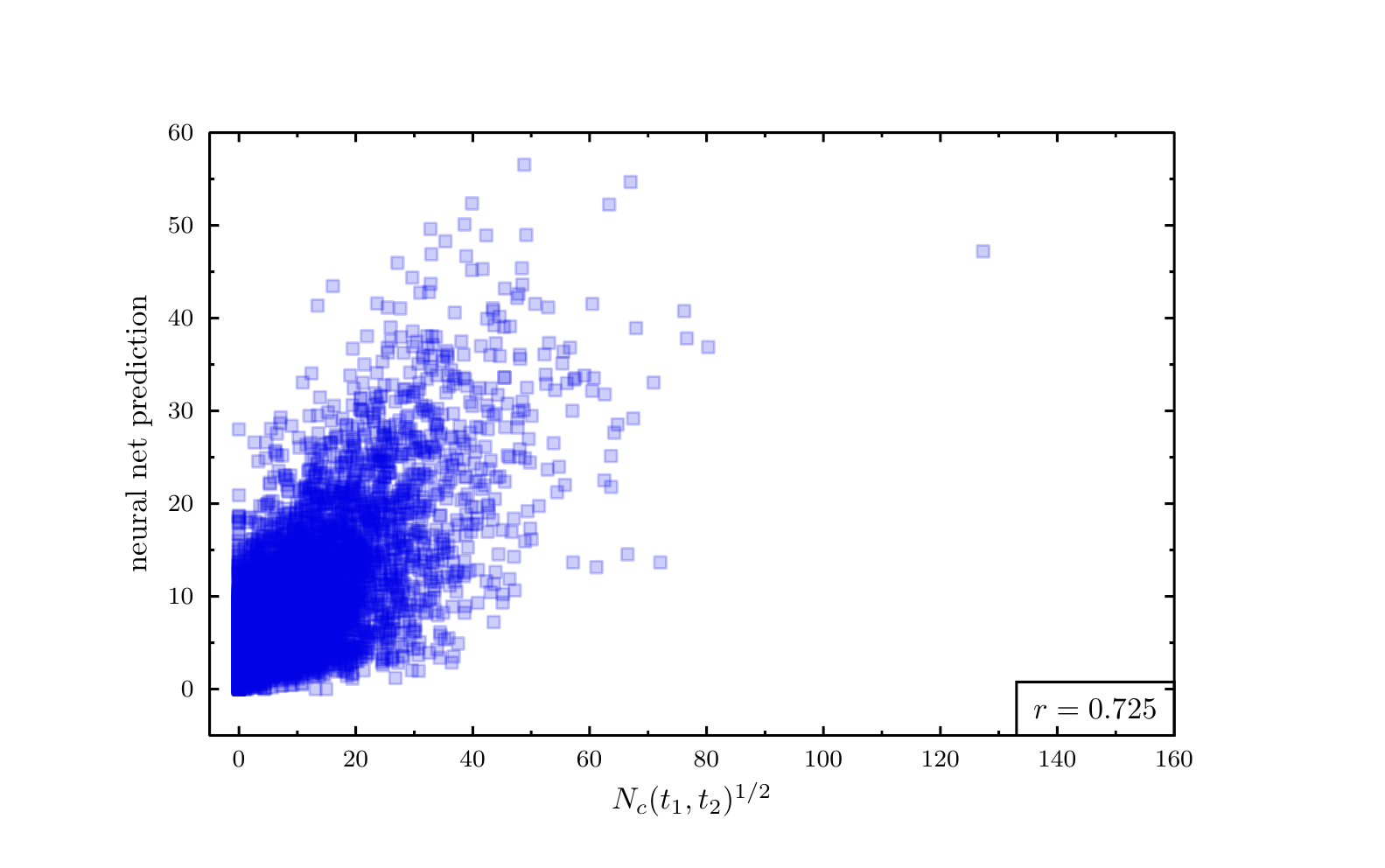}
 \caption{The neural network's prediction for $ N_{\textrm{c}}(t_1, t_2)^{1/2} $ compared to the actual value for all authors in the validation dataset. Shown is the result from the first round of cross-validation.}
 \label{fig:net}
\end{figure}

The correlation coefficients $ r $ betweeen the neural network's prediction and $ N_\textrm{c}(t_1, t_2)^{1/2} $ is $ r = 0.725 \pm 0.007 $ (see Fig.~\ref{fig:net}).
The error here and in the following is the standard deviation across the 20 rounds of cross-validation.
In comparison, the naive $h$-index predictor described above yields $ r = 0.551\pm0.008 $.
We clearly see that the neural network performs significantly better than the naive $h$-index predictor.

Next, we tested whether any one type of input data is especially important to the performance of the neural network
by training the network with single types of input data separately removed. This, we found, barely
changes the results. The  biggest impact comes from removing the paper dates and the number of citations,
which make the correlation coefficient drop to $ r = 0.684\pm0.007 $ and $ r=0.718\pm0.005$, respectively, which is still very close to the original $r=0.725\pm0.007$. We may speculate that the network gathers its information from combinations of input which
are themselves partly redundant, so that removing any single one has little effect.

Finally, we checked whether the neural net still performs better than the $h$-index when given only citation counts as input data. This resulted in a correlation coefficient of $r=0.579\pm0.007 $.
We see that the neural net performs better than the $h$-index even with only the number of citations as input data.

For more details on the comparison between our results and those of Ref.~\cite{Hirsch2007}, please see Appendix C.

\subsection{Comparison with earlier machine learning predictions}
\label{second}

\begin{figure}[t]
 \centering

 \includegraphics[width=.9\textwidth]{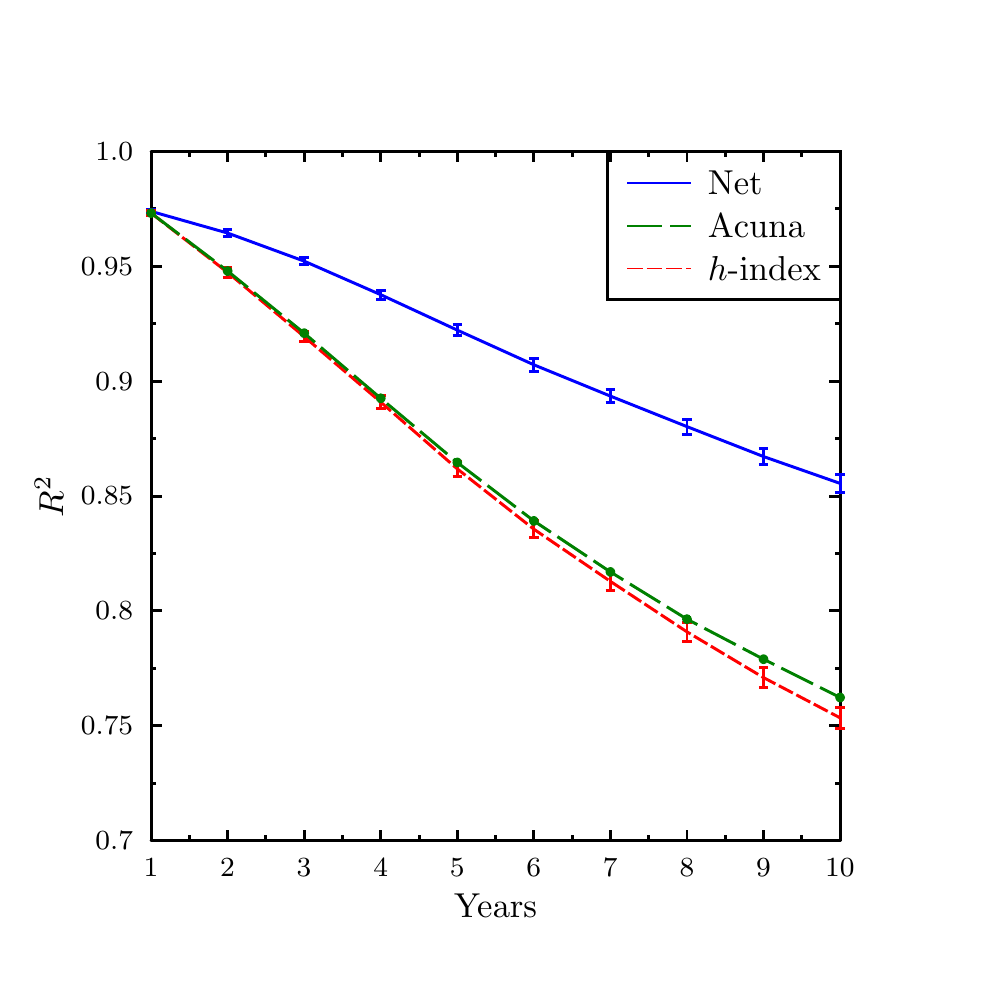}
 \caption{$ R^2 $ of the prediction of the cumulative $ h $-index as a function of years after cutoff.}
 \label{fig:time}
\end{figure}

For the second approach, we compare the predicted with the actual cumulative $ h $-index for $n=1,2 \dots 10 $ years after the
cutoff date. We quantify the goodness of this prediction with the coefficient of determination, $ R^2 $,
 both for the neural network and for the naive $ h $-index predictor discussed at the beginning of this section.
Note that we calculate $ R^2 $ separately for each $ n $ by restricting the predictions as well as the actual values to that particular $ n $.
The result is shown in Fig.~\ref{fig:time}.
The error bars show $\pm 1 $ standard deviations calculated from the 20 rounds of cross-validation.

We see that the neural net and the naive $ h $-index predictor are similarly predictive for $ n = 1 $, but the neural network's prediction becomes better in comparison for larger $ n $. This agrees with the findings Refs.~\cite{Acuna2012,7991559}.

\begin{figure}[t]
 \centering
 \includegraphics[width=.49\textwidth]{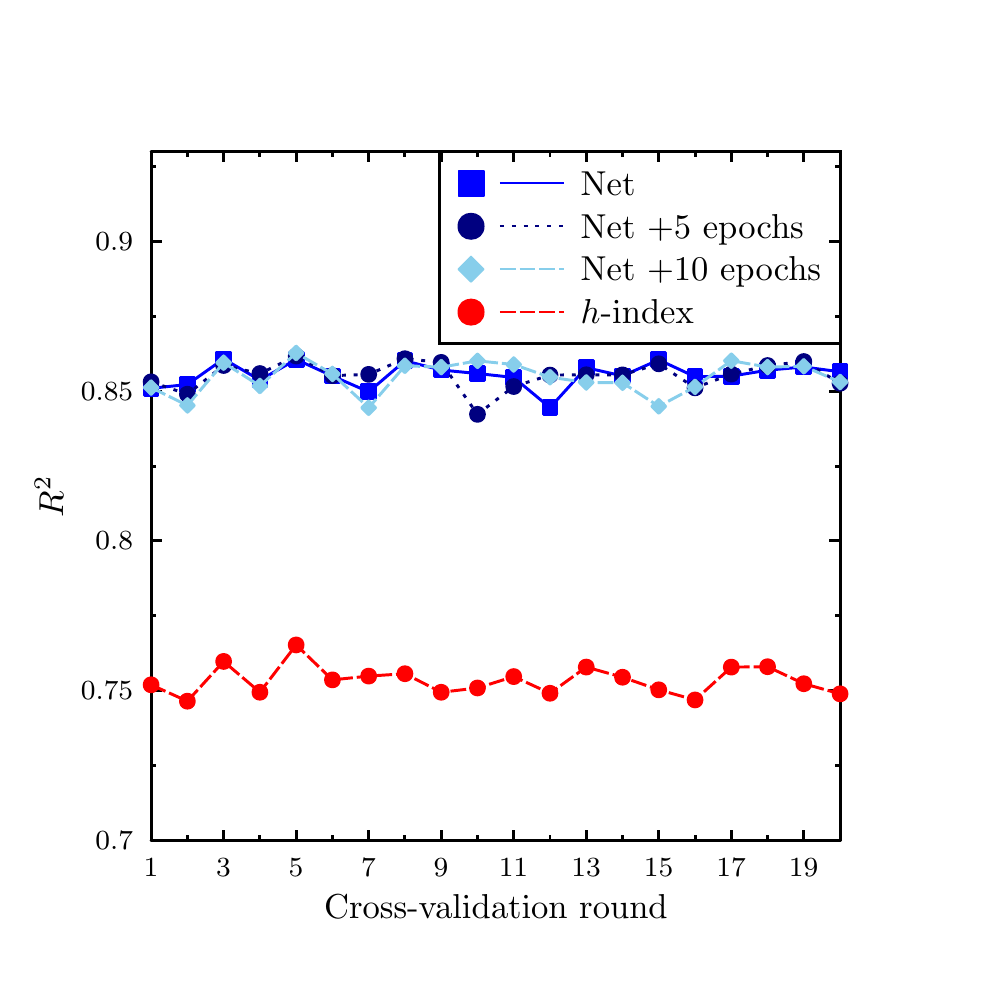} \includegraphics[width=.49\textwidth]{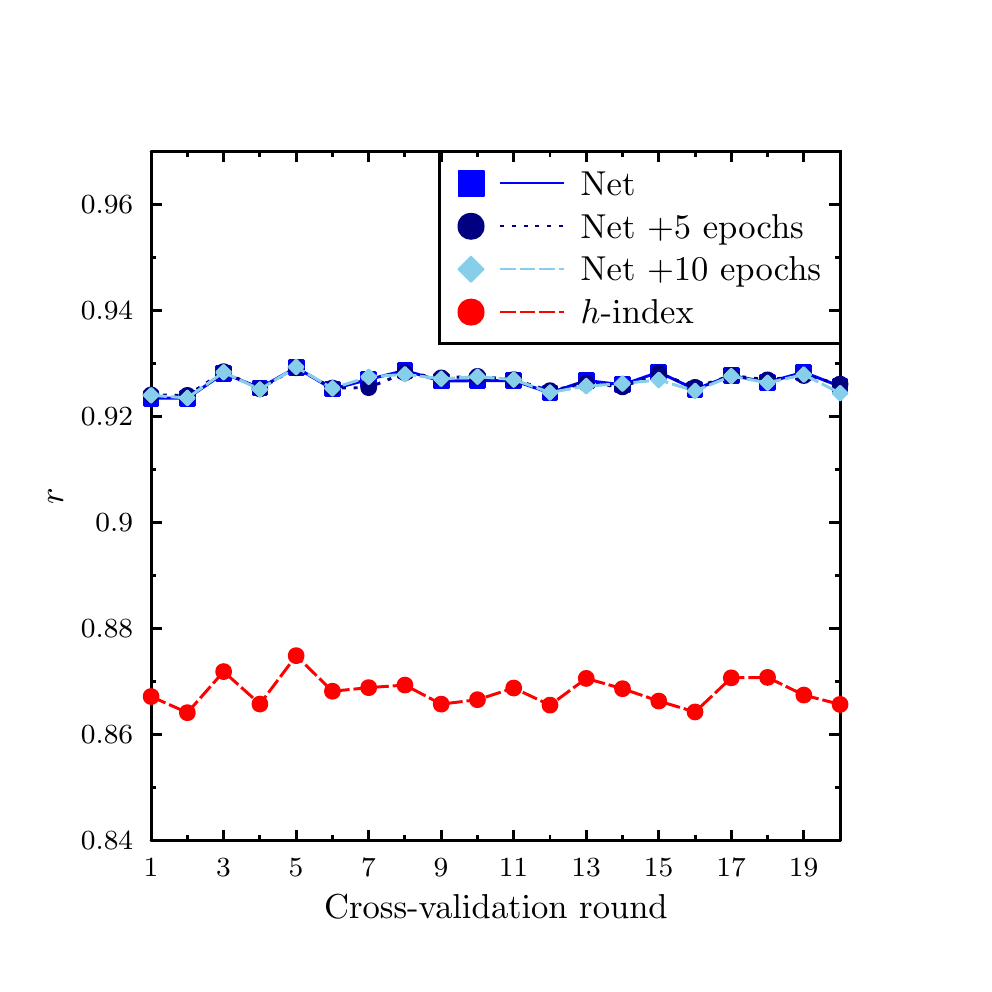}

 \caption{Left: Neural network's and naive $h $-index predictor's $R^2$ after $ n = 10 $ years for each round of cross-validation. Right: Neural network's and naive $h $-index predictor's $r$ after $ n = 10 $ years for each round of cross-validation.}
 \label{fig:cv-fluctuations}
\end{figure}

As mentioned in section~\ref{aim}, there are fluctuations of the neural network's performance with the rounds of cross-validation.
These fluctuations are illustrated in Fig.~\ref{fig:cv-fluctuations}.
In particular, Fig.~\ref{fig:cv-fluctuations} shows the neural network's $ R^2 $ (Fig.~\ref{fig:cv-fluctuations}, left) and $ r $ (Fig.~\ref{fig:cv-fluctuations}, right) after $ n = 10 $ years for the different rounds of cross-validation.
We see that there are non-negligible fluctuations in the network's performance.
A possible explanation could be that the fluctuations are due to the neural network overfitting on particular kinds of splits of the whole dataset into training and validation data.
However, we think it is more likely that the fluctuations are due to intrinsic properties of the dataset.
This is because the naive $ h $-index predictor has only a few tens of parameters so that overfitting should not be an issue but Fig.~\ref{fig:cv-fluctuations} shows that there are comparable fluctuations in this naive $ h $-index predictor's performance nonetheless.

\begin{figure}[t]
 \centering
 \includegraphics[width=.49\textwidth]{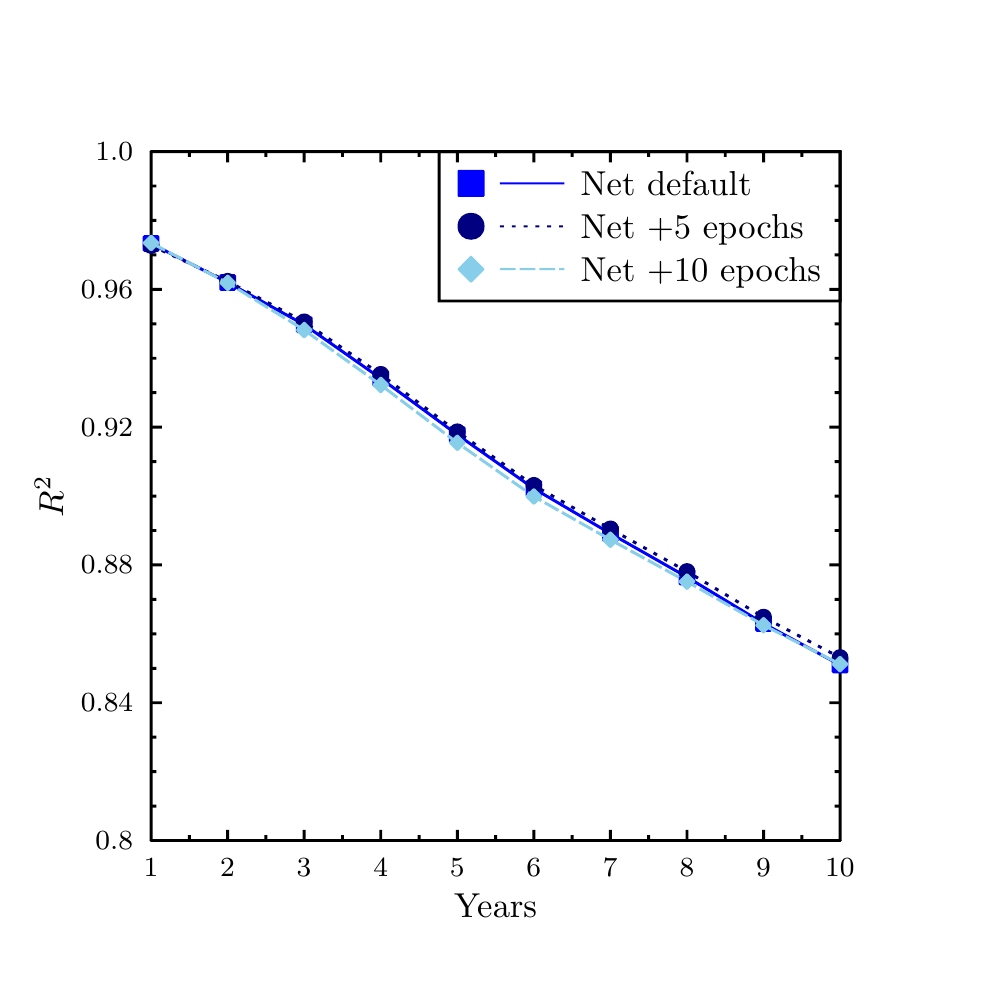} \includegraphics[width=.49\textwidth]{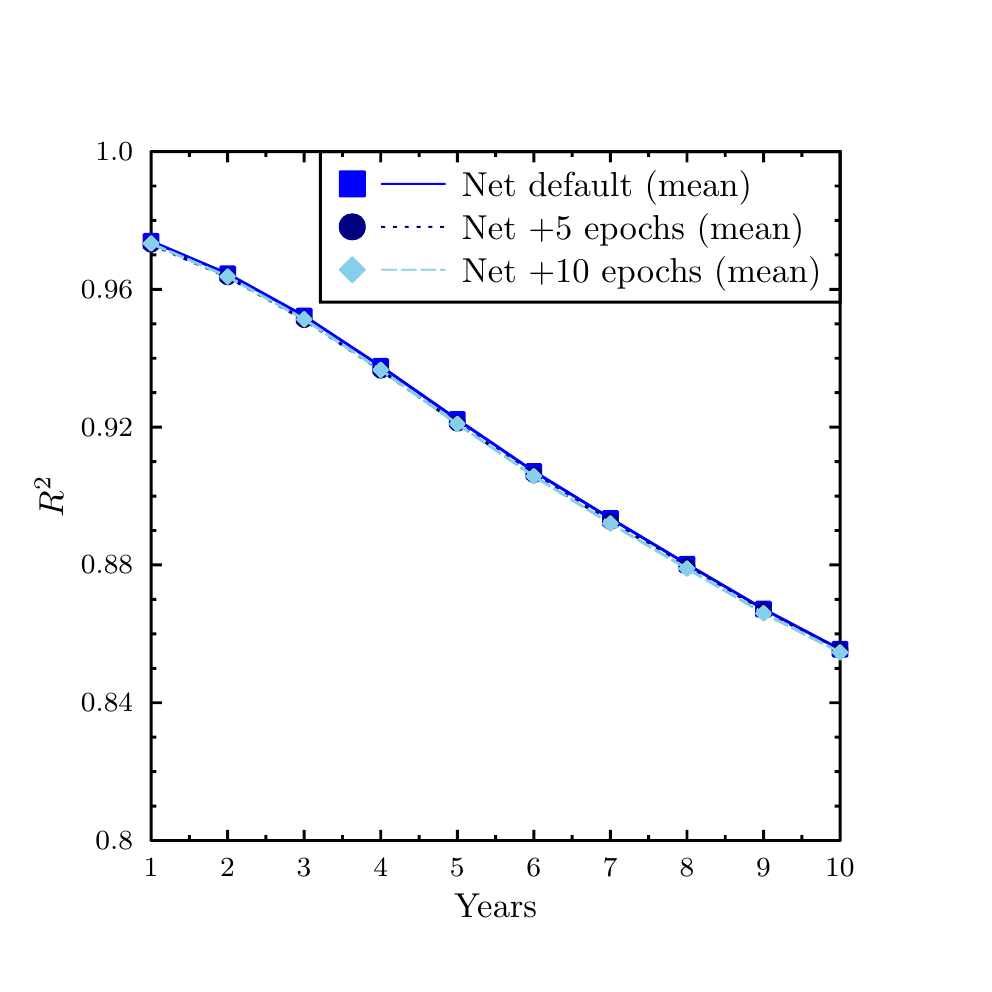}

 \caption{Left: Network trained
by our default value of 150 epochs, compared with training for 155 and 160 epochs for the first round of cross-validation. Right: Average across 20 rounds of cross-validation of the network trained by our default value of 150 epochs, compared with training for 155 and 160 epochs.}
 \label{fig:epochs-fluctuations}
\end{figure}

In Fig.~\ref{fig:epochs-fluctuations}, left, we show the result for the neural network operating on the training and validation datasets of the first round of cross-validation when training for 5 and 10 additional epochs.
We see that there are fluctuations of the neural network's performance with the training epoch which typically affect $ R^2 $ at or below the
one percent level.
In Fig.~\ref{fig:epochs-fluctuations}, right, we show the result of averaging across all round of cross-validation, where the averaging is done separately for the neural networks obtained after training for 150, 155, and 160 epochs.
We see that the fluctuations with the training epoch have cancelled and the difference between the averages after training for 150, 155, and 160 epochs is negligible.

Quantitatively, our results give higher values of $ R^2 $ than both Ref.~\cite{Acuna2012} (0.48 after 10 years) and Ref.~\cite{7991559} (0.72 after 10 years).
However, since not only our neural network but also our simple $ h $-index predictor give higher $ R^2 $-values than Refs.~\cite{Acuna2012} and \cite{7991559}, this difference is probably partly due to the different datasets. We have therefore also applied the predictor
proposed in \cite{Acuna2012} to our data-set, with the results shown in Figure \ref{fig:time} (see Appendix D for details). The predictability of our data-set is
indeed higher than that studied in \cite{Acuna2012}, but the prediction of our network still outperforms the previous study.
Unfortunately, a similar direct comparison to the results from \cite{7991559} which used yet another data-set is not possible. Still,
our value of $R^2=0.856\pm0.004$ after ten years is remarkably predictive, especially given that the methods we
have employed here are likely to improve further in the soon future.

\begin{figure}[ht]
 \centering
 \includegraphics[width=.49\textwidth]{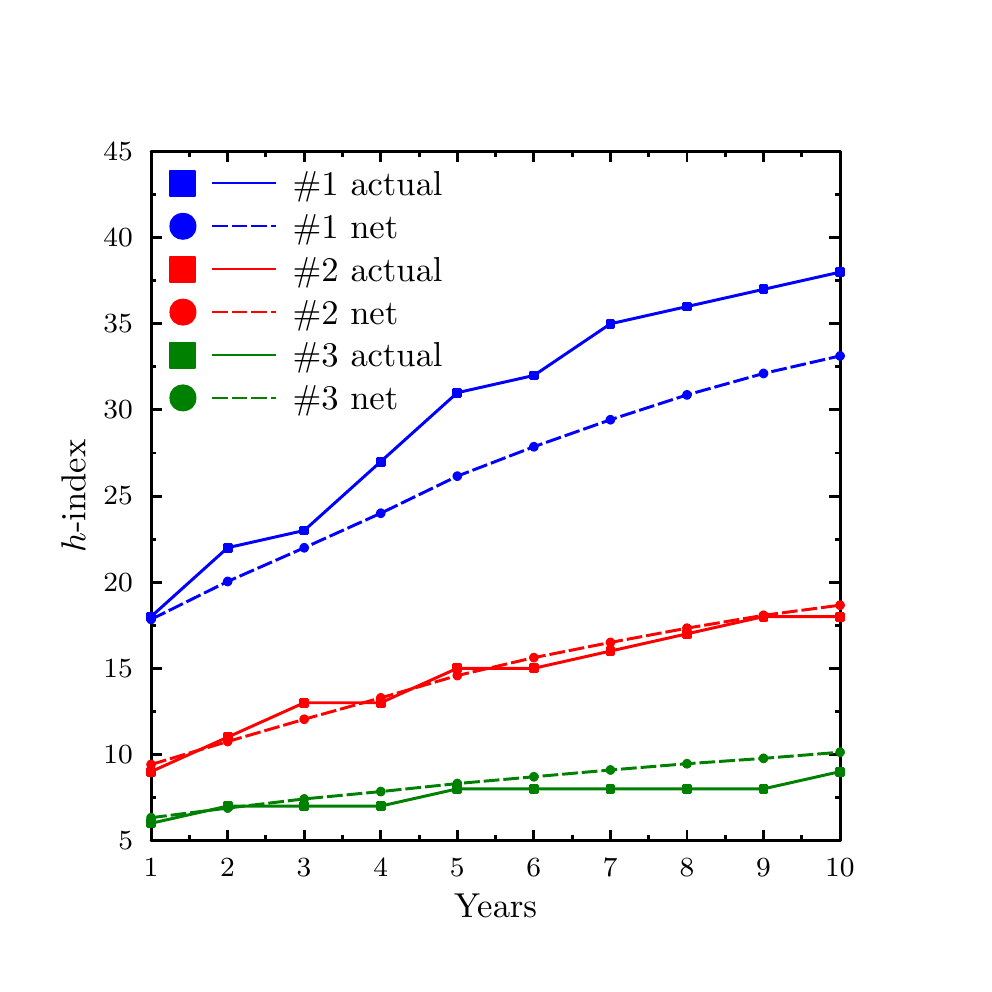}
\includegraphics[width=.49\textwidth]{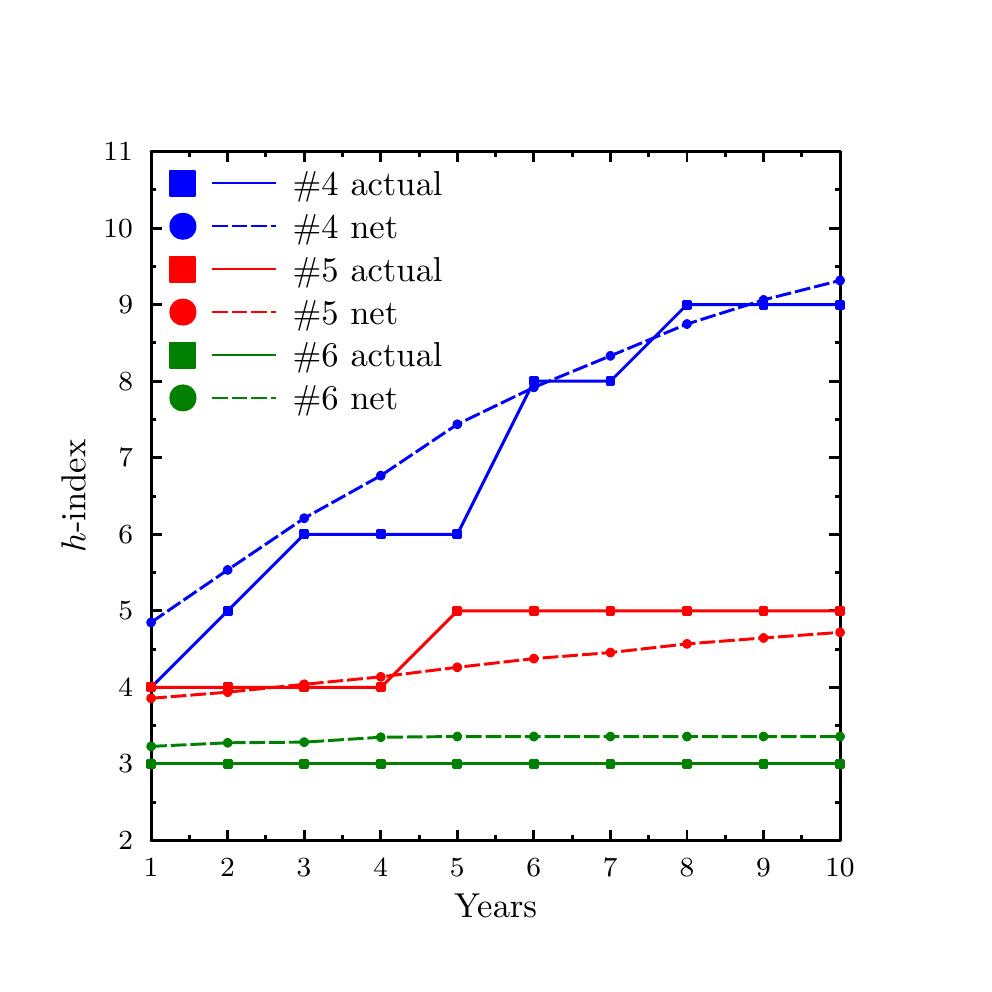}
 \caption{Example trajectories for actual development of the $h$-index over time (solid/squares) and trajectories predicted by the network (dashed, circles) for the training and validation data of the first round of cross-validation.}
 \label{fig:expl}
\end{figure}

Since the value of $R^2$ by itself is not so illuminating, we show in Figure \ref{fig:expl} some examples for which
we display the actual $h$-index versus the network-prediction with the training and validation datasets from the first round of cross-validation.\footnote{The reader be warned that these examples
were not randomly chosen. We hand-selected a set of noticeably different $h$-index outcomes for
purely illustrative purposes.}

\begin{figure}[ht]
 \centering
 \includegraphics[width=.49\textwidth]{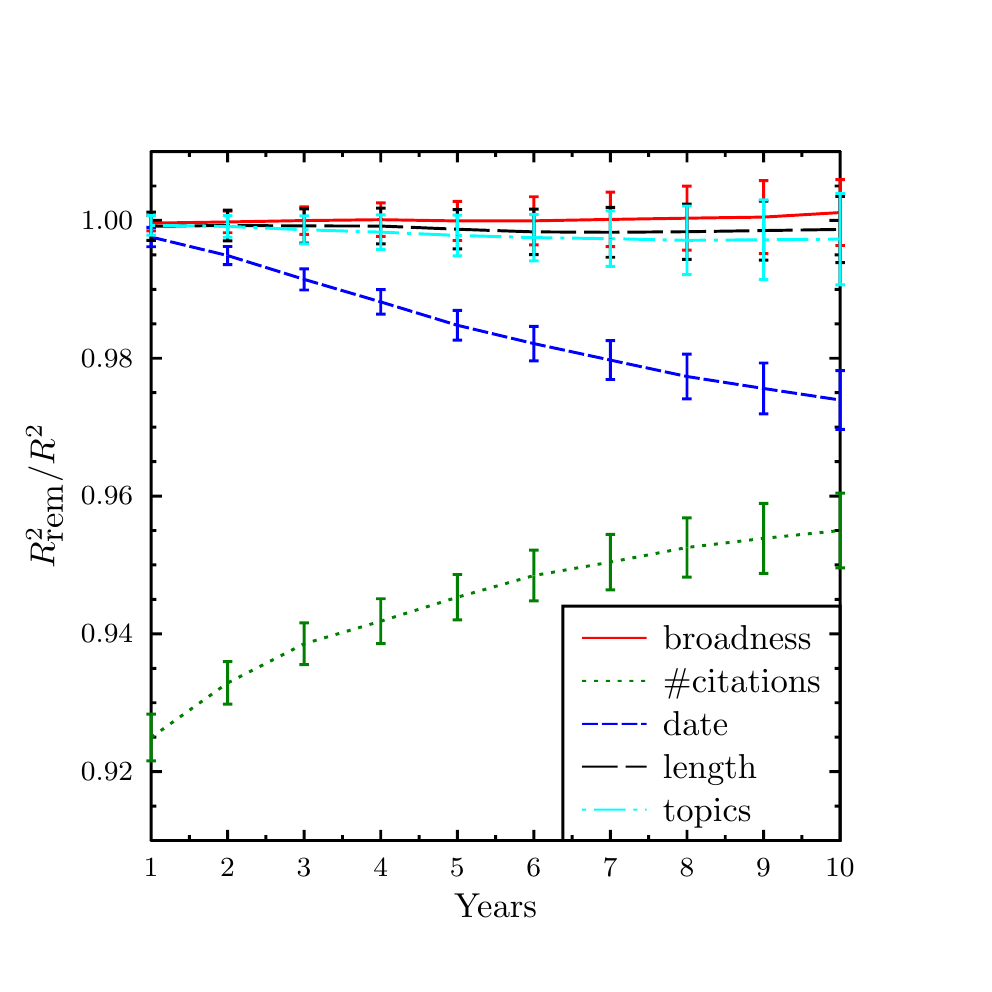}
\includegraphics[width=.49\textwidth]{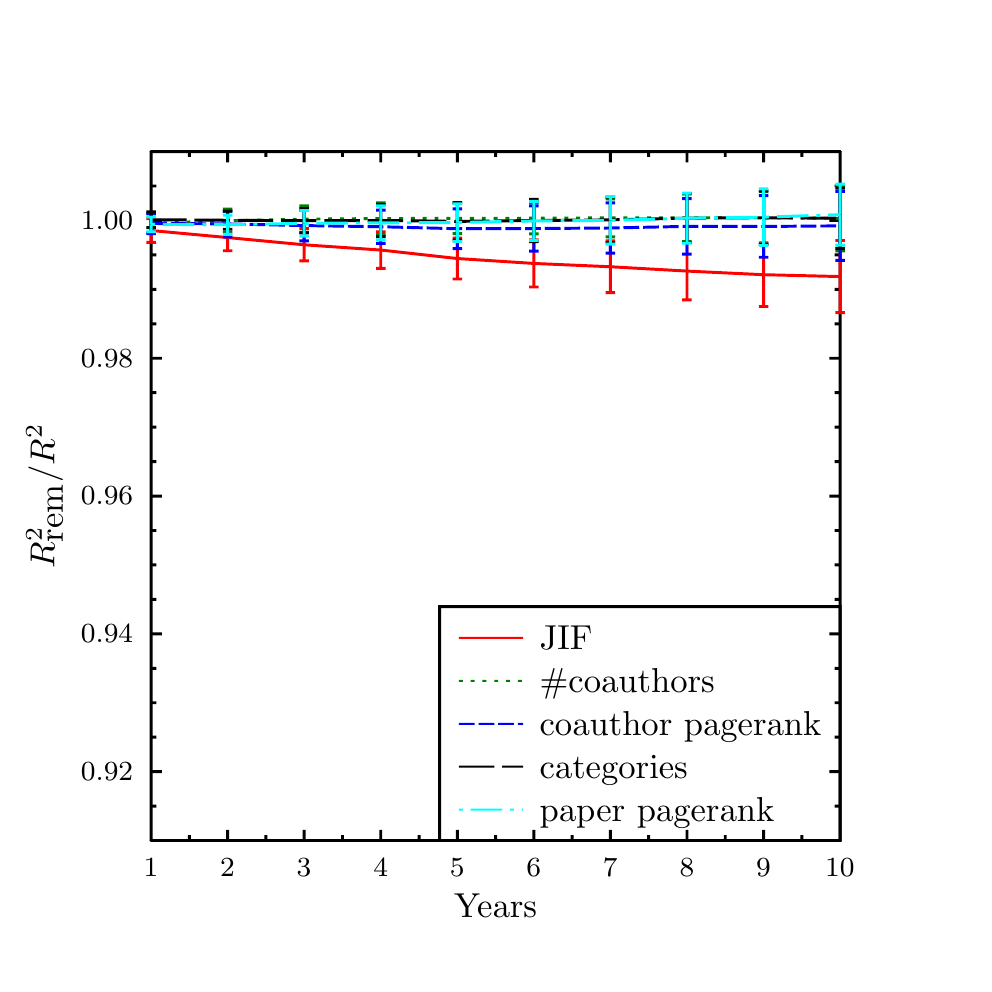}
 \caption{Ratio of $ R^2 $ of the neural net to the same indicator for the neural net with various input data removed, $R_{\rm rem}^2$, as a function of years past cutoff.}
 \label{fig:timeremoved}
\end{figure}

Here too we have investigated how important various input data are to the neural network's performance by removing
one at a time. The results for $ n = 1, 5, $ and $ 10 $ are shown in Fig.~\ref{fig:timeremoved}, where
we plot the ratio of the coefficients of determination with all input ($R^2$) and with certain inputs
removed ($R_{\rm rem}^2$). The lower the ratio, the more important the removed input was.
The error bars show $\pm 1 $ standard deviations calculated from the 20 rounds of cross-validation.

We see that for $ n = 1 $, the number of citations is the only important input, while for $ n > 1 $ other inputs gain importance with the number of citations still being the most important one.
That the citation data are most important for $ n = 1 $ agrees with the results of Ref.~\cite{Acuna2012} and Ref.~\cite{7991559}. We also see from Fig.~\ref{fig:timeremoved} that $ R^2_\textrm{rem}/R^2 $ is always larger than $ 0.9 $ which again
indicates that the input data are partly redundant. However, the changes we notice due to some
input removals are so small that fluctuations in the
training results are no longer negligible.

\section{Discussion}
\label{conc}

We have demonstrated here that neural nets are powerful tools to make
predictions for the citations a
researcher's work accumulates. These predictions are likely to improve
in the future. One of the major limitations of this present study, for example,
is that our sample does not include papers which are not on the arXiv,
and that about one of five published papers could not be associated with
a Journal Impact Factor (see Appendix A).
But the more bibliometric data becomes available the more input the
network can be fed, and thus predictivity is bound to become better for
some more time.

The methods we used here are straight-forward to implement and do not require
much computing resources. It is thus foreseeable that in the near
future the use of neural nets to predict researcher's promise will become
more widely spread.

We therefore want to urge the community to not ignore this
trend in the hope it will go away. It would benefit
academic research if scientists themselves proposed a variety of
predictors, and offered a variety of data, to more accurately present the variety
of ways to do high-quality research.

In this work we focused on the $h$-index in order to compare
our results with previous results, and also because this value is
easy to extract from existing data. But a lot of valuable
information about researchers and their work is presently difficult or impossible to
obtain and analyze. For example,
how often researchers are named in acknowledgements, their
seminar and conference activity, the frequency by which they act as peer reviewers
and/or editors, or how specialized their research topic is. All these are important
factuals about the many individual approaches to research. Providing and analyzing
such data would enable us to develop measures for success tailored
to specific purposes and thereby avoid that researchers focus efforts
on optimizing citation counts.

\section*{Acknowledgements}

TM thanks Sebastian Weichwald for helpful discussions.
This work was supported by the Foundational Questions Institute (FQXi).

\section*{Appendix A}

The building blocks of most of the input data for the neural network are lists where the position in the list corresponds to a paper and the value at a certain position corresponds to some input data associated with the corresponding paper.
E.g. there is a list containing the number of citations for each paper of the given author which in Python syntax would look like $ [ 130, 57, ... ] $, meaning the most-cited paper of this author has $ 130 $ citations, the second-most cited paper has $ 57 $ citations etc.
A corresponding list for the number of coauthors would look like $ [ 0, 1, ... ] $, meaning the paper with $ 130 $ citations was a single-authored paper, the paper with $ 57 $ citations has two authors etc.

Since different authors have different numbers of papers, these lists will have different lengths for different authors.
However, our neural network requires a fixed-size input.
Therefore, we take the lists for all authors to have the same length as those of the author with the largest number of papers.
The positions in the lists that do not correspond to a paper, are filled with zeros.

The biggest part of the input to the neural network is then the list of all the lists described above and consists of
\begin{enumerate}
 \item A list which contains 1 at each position in the list which corresponds to a paper and 0 at each position which corresponds to zero-padding.
 \item A list which contains the number of citations of each paper.
 \item A list which contains the publication date of each paper relative to the cutoff date.
 \item A list which contains each paper's pagerank \cite{pagerankimpl}, an interactive measure of relevance that works similar to Google's pagerank algorithm
just that, instead being based on hyperlinks, a paper's pagerank is based on the  citation graph.
The pagerank is calculated from the citation graph at the cutoff date, 1/1/2008.
 \item A list which contains each paper's length.
 \item A list which contains 0 for each paper with an empty journal reference and 1 for each paper with a non-empty journal reference.
 \item A list which contains the {\sc JIF} of the journal each paper is published in (further details below). If a paper is not published or no {\sc JIF} is known for a journal, we take the corresponding input to be 0.
The {\sc JIF}s are taken at the cutoff date.
 \item A list which contains the number of coauthors of each paper.
 \item Three lists which contain the coauthors' minimum, maximum, and average pagerank.
 Here, the pagerank is calculated from the coauthor graph at the cutoff date.
 \item For each arXiv category a list which contains zeros except at position which correspond to papers which are in the respective category.
 In order to reduce the amount of data, categories of the form a.b are all treated as category a, e.g. astro-ph.CO and astro-ph.HE are treated as the same category.
 \item For each paper, a 50-dimensional vector representing a paper's latent topic distribution, obtained from the keyword analysis done by \cite{Tom} when operating on the arXiv data up to the cutoff.
\end{enumerate}

The final input to the neural network is a `broadness' value calculated from a keyword analysis \cite{Tom} with the same data as the paper topics described above. This broadness quantifies how widely spread the topics which an author publishes on are over all arXiv categories.
Since this broadness value is one value per author and not one value per paper it is handled separately from the other inputs to the neural network.

All input data except the categories and the paper vectors is normalized to unit variance and zero mean.
More specifically, for each input (e.g. number of citations, publication date, broadness, ...) a transformation is determined from the training data such that this transformation brings the given input to zero mean and unit variance for the training data across all authors.
This transformation is then applied both to the training and the validation input data.

To assign the {\sc JIF}s, we associate papers in our database with a journal by heuristically matching the journal reference given in the arXiv metadata to the journal abbreviation from Ref.~\cite{JCR2017}.
Concretely, we reduce both the journal reference in the arXiv metadata and the journal abbreviation from Ref.~\cite{JCR2017}  to  lower-case alphanumeric characters and cut at the first numeric character.
Next, we remove the suffixes 'vol' and 'volume'  if present.
If the two values obtained this way are identical, we consider the given arXiv paper to be published in the corresponding journal.

To reduce the number of papers where this procedure does not work, we have further used a manually assembled translation table.
This table contains identifications between reduced arXiv journal references to journals from Ref.~\cite{JCR2017} for
which the method outlined in the previous paragraph does not work. The table allows us to match the 69 most frequent reduced journal references from the arXiv that
could not be mapped by the previous method.
By this procedure, we have assigned a Journal Impact Factor to 378,134 of the 477,176 papers with a non-empty arXiv journal reference.

\section*{Appendix B}

\begin{lstlisting}[
    language=Python,
  %  caption={Implementation of the neural network architecture using Keras},
    label={lst:net}]

from keras.models import Model
from keras.layers import Input, Dense, Conv1D, concatenate, \
    GlobalAveragePooling1D

perpaper_inputs = Input(shape=perpaper_shape,
                        name='perpaper_inputs')
perauthor_inputs = Input(shape=perauthor_shape,
                         name='perauthor_inputs')

tmp = Conv1D(
    filters=70,
    kernel_size=1,
    strides=1,
    activation=activation,
    input_shape=perpaper_shape)(perpaper_inputs)

tmp = GlobalAveragePooling1D()(tmp)

tmp = concatenate([tmp, perauthor_inputs])
tmp = Dense(units=70, activation=activation)(tmp)

outputs = Dense(units=10, activation='relu')(tmp)

model = Model(inputs=[perpaper_inputs, perauthor_inputs],
              outputs=outputs)
\end{lstlisting}

\section*{Appendix C}

\begin{figure}[h]
 \centering
\vspace*{-.5cm}
\hspace*{-0.5cm} \includegraphics[width=1.1\textwidth]{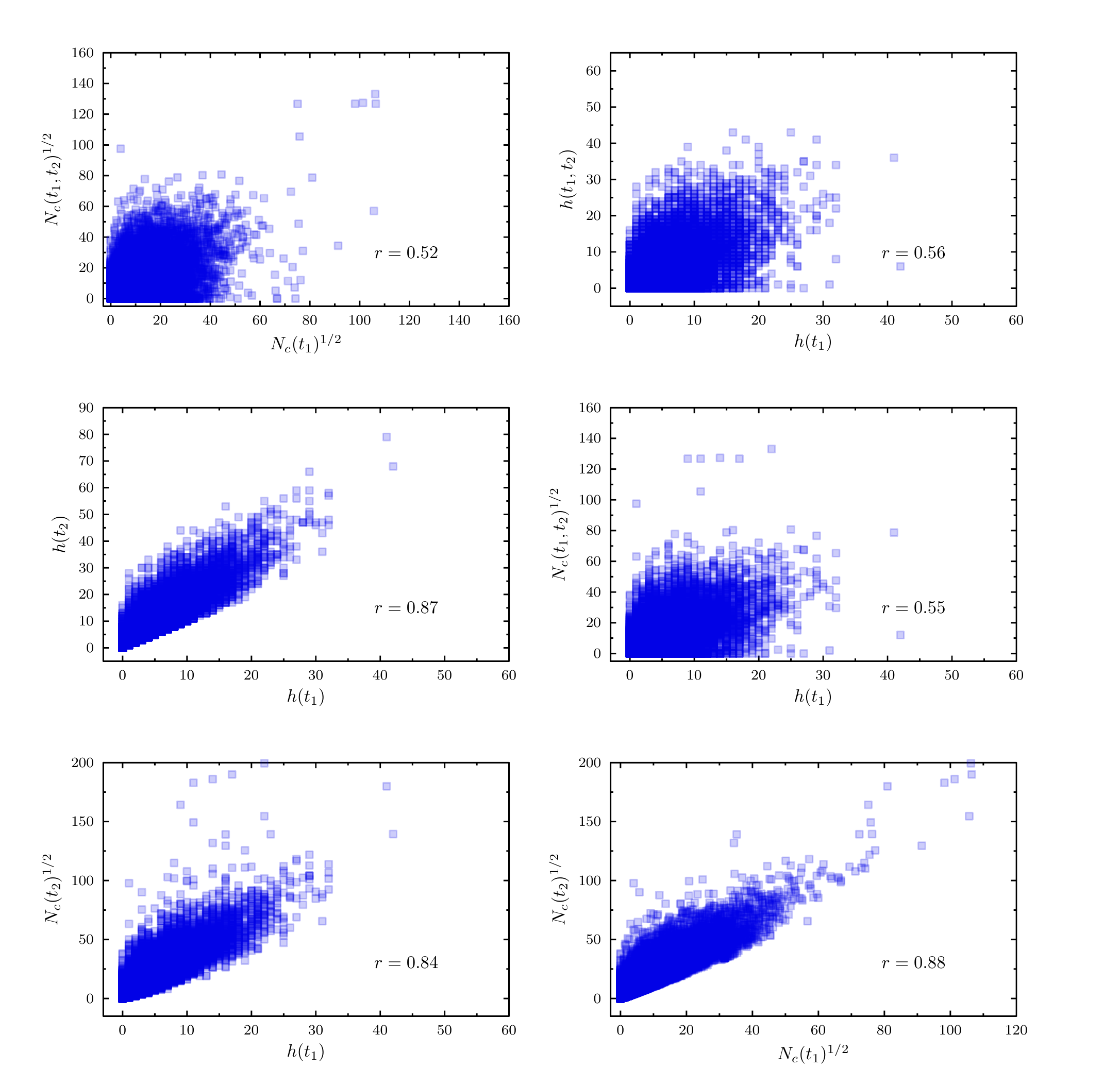}
 \caption{Correlation of various quantities calculated from the complete dataset including both training and validation data with the respective correlation coefficient $ r $.
 The notation is that of Ref.~\cite{Hirsch2007}.
 E.g., $ h(t_1) $ is the $ h $-index as calculated from an author's first 10 years of publishing, $ h(t_2) $ is the cumulative $ h $-index after an author's first 20 years of publishing, and $ h(t_1, t_2) $ is the $ h $-index calculated from the second 10 years of an author's publishing excluding papers written and citations received outside this period of time.
 For different datasets, these plots can be found in Ref.~\cite{Hirsch2007}.}
 \label{fig:hirsch}
\end{figure}

The correlation coefficient obtained from the naive $ h $-index predictor is roughly the same as the correlation coefficient obtained by plotting $ N_{\textrm{c}}(t_1, t_2)^{1/2} $ over the $ h $-index at the time of the cutoff for both the training and validation data and calculating the correlation coefficient from that, see Fig.~\ref{fig:hirsch}.
Note that this second way of calculating a correlation coefficient corresponds to what was done in Ref.~\cite{Hirsch2007}.

The correlation coefficients from Fig.~\ref{fig:hirsch} are consistenly smaller than those of the sample {\sc PRB80} from Ref.~\cite{Hirsch2007} but higher than those of the sample {\sc APS95} from Ref.~\cite{Hirsch2007}.
See Ref.~\cite{Hirsch2007} for a discussion of the differences between the samples {\sc PRB80} and {\sc APS95} regarding their differing correlation coefficients.
Our sample differs from both {\sc PRB80} and {\sc APS95} in both the data source and the cuts applied.
Therefore, it is not surprising that there are differences in the resulting correlation coefficients and the results
are not directly comparable.
One important difference is that we employ the same cutoff, 1/1/2008, for all authors while Ref.~\cite{Hirsch2007} applies a different cutoff for each author at 12 years after each author's first paper.

\section*{Appendix D}

We used what the authors of \cite{Acuna2012} refer to as the ``simplified model,'' that -- as they have
shown -- performs almost as well as their full model on their data-set. We
changed the selected journals from Nature, Science, Nature Neuroscience, PNAS and Neuron to Science, Nature, PNAS, and PRL.
As laid out in the supplementary material of \cite{Acuna2012}, we used the R-package `glmnet' with
$\alpha = 0.2$.
Note that we did not employ our own Monte Carlo cross-validation here. Instead -- as was done in \cite{Acuna2012} -- we relied on the cross-validation included in the `glmnet` package.

\bibliography{net-notes_s}
\bibliographystyle{unsrt}

\end{document}